\documentclass[letterpaper]{article} 
\usepackage{aaai24}  
\usepackage{times}  
\usepackage{helvet}  
\usepackage{courier}  
\usepackage[hyphens]{url}  
\usepackage{graphicx} 
\usepackage{natbib}  
\usepackage{caption} 
\usepackage{algorithm}
\usepackage{algorithmic}

\usepackage{xcolor}

\usepackage{newfloat}
\usepackage{listings}
\DeclareCaptionStyle{ruled}{labelfont=normalfont,labelsep=colon,strut=off} 
\floatstyle{ruled}
\newfloat{listing}{tb}{lst}{}
\floatname{listing}{Listing}
\nocopyright 

\title{The Contribution of XAI for the Safe Development and Certification of AI: \newline An Expert-Based Analysis}
\author {
    Benjamin Fresz \textsuperscript{\rm 1,2},
    Vincent Philipp Göbels\textsuperscript{\rm 3},
    Safa Omri\textsuperscript{\rm 3},
    Danilo Brajovic\textsuperscript{\rm 1,2},\\
    Andreas Aichele\textsuperscript{\rm 1,2},
    Janika Kutz\textsuperscript{\rm 3},
    Jens Neuhüttler\textsuperscript{\rm 3},
    Marco F. Huber\textsuperscript{\rm 1,2}
}
\affiliations {
    \textsuperscript{\rm 1}Fraunhofer Institute for Manufacturing Engineering and Automation IPA, Stuttgart, Germany\\
    \textsuperscript{\rm 2}Institute of Industrial Manufacturing and Management IFF, University of Stuttgart, Germany\\
    \textsuperscript{\rm 3}Fraunhofer Institute for Industrial Engineering IAO, Heilbronn, Germany\\
    \{benjamin.fresz, danilo.brajovic, andreas.aichele\}@ipa.fraunhofer.de, \\
    \{vincent-philipp.goebels, safa.omri, janika.kutz, jens.neuhuettler\}@iao.fraunhofer.de,
    marco.huber@ieee.org
}

\begin{document}

\maketitle

\begin{abstract}
Developing and certifying safe---or so-called trustworthy---AI has become an increasingly salient issue, especially in light of upcoming regulation such as the EU AI Act.
In this context, the black-box nature of machine learning models limits the use of conventional avenues of approach towards certifying complex technical systems.
As a potential solution, methods to give insights into this black-box---devised in the field of eXplainable AI (XAI)---could be used. 
In this study, the potential and shortcomings of such methods for the purpose of safe AI development and certification are discussed in $15$ qualitative interviews with experts out of the areas of (X)AI and certification.
We find that XAI methods can be a helpful asset for safe AI development, as they can show biases and failures of ML-models, but since certification relies on comprehensive and correct information about technical systems, their impact is expected to be limited.
\end{abstract}

\section{Introduction}



In the rapidly evolving domain of machine learning (ML), the integration of ML systems into safety-critical applications presents unique challenges, primarily due to the ML-inherent opacity. Often characterized as ``black-box" systems, such models are based on learning patterns instead of being explicitly programmed, thus complicating transparency and reliability \cite{castelvecchi2016}. This opacity not only challenges their integration into environments where safety is paramount but also impedes established system certification processes.

The emerging field of eXplainable Artificial Intelligence (XAI) seeks to address this challenge by improving the transparency of ML models \cite{rai2020}. XAI aims to make the decision-making processes of AI systems comprehensible to human stakeholders, thereby potentially increasing their trustworthiness and facilitating their integration into regulated domains \cite{Martinie.2021, Brajovic.2023}. 
Despite the growing research interest in XAI, its practical utility in enhancing the safety and certification of AI systems has not been thoroughly investigated.

This paper addresses this gap by examining the potential of XAI in the certification of AI systems. 
It discusses whether the current state of XAI tools can be integrated into certification processes and evaluates their practical utility through the experiences of practitioners. Specifically, this research addresses three primary questions:
\begin{enumerate}
    \item Does XAI function as a debugging tool in practice, and what implications does this have for the safety of AI systems?
    \item Is it feasible to incorporate XAI into existing and future certification frameworks for AI systems?
    \item What are the positive and negative experiences of practitioners using XAI in the field?
\end{enumerate}
To answer these questions, we conducted qualitative interviews with $15$ experts who operate at the intersection of AI development and certification. 
Furthermore, we make a distinction between XAI’s role as an auxiliary tool for debugging, which can improve ML model development, and its more consequential role as a certification instrument that could offer assurances about an ML model’s functionality.
This paper aims to expand the discourse on AI safety by critically analyzing the role of XAI within the certification landscape, assessing both its potential and its limitations.
To the best of our knowledge, this paper offers the first in-depth exploration of how XAI can be utilized in the certification and safeguarding of AI systems.





The paper is organized as follows: After an introduction into the safe development of technical systems, the related works are presented, with a focus on the context of certification of AI systems.
In Section~3, readers are provided with the necessary background knowledge of XAI and related techniques.
In Section~4, the methodology and participant profiles are introduced, followed by the presentation of the interview results in Section~5.
Further pathways for XAI and limitations of the used approach are discussed in Section~6. The paper closes with a summary in Section~7.

\subsection{Safe Development of Technical Systems}


For non-AI products, safe development and certification processes are well established, as they are subject to numerous legal and standardized requirements. For example, the Machinery Directive 2006/42/EC of the European Union regulates the provisions for placing machinery on the market in the European Economic Area.
 A key point of this directive is the minimum requirements for safety and health protection. Specific requirements are derived from references to corresponding harmonized standards.
 For technical systems with AI functionalities, which are the focus of this paper, the area of electrical, electronic and programmable electronic systems is most likely to apply.
 If a system from this area is developed with a safety function, IEC 61508 or one of its sister standards for specific areas of application (e.g., ISO 26262 for automobiles) applies.
 These standards describe a process model, methods to be used, and various required activities and work products.
 The basic procedure is to identify potential situations that pose a risk to life and limb. The relevance or dangerousness of situations is determined by means of a risk assessment.
 In the case of particularly dangerous situations, further methods must be used to avoid systematic errors.
 In the case of random faults, a quantitative assessment of the components with a maximum permissible probability of failure is required.
 Companies and/or products are certified to confirm compliance with these requirements.
 An independent body checks whether the requirements specified in the standard have been met.

\section{Related Works}
Although the certification of AI systems is not standardized as of now, multiple scientific publications provide potential avenues of approach.
Some of them are presented in the following, to give an overview over the challenges and potential solutions for safe AI development and certification.

\subsection{AI Certification Challenges}



The certification of AI systems is a complex and evolving challenge, as highlighted by multiple publications \cite{falcini2017, stoica2017, vanderlinde2022, mahilraj2023}.
They acknowledge that certifying AI Systems is particularly difficult due to several factors that diverge from those encountered in traditional software certification. 
Because of this, \citet{falcini2017} specifically address the need for new certification schemes in the automotive industry, while \citet{stoica2017} emphasize the importance of AI systems that can make safe and timely decisions in unpredictable environments.
Multiple publications discuss the need for new approaches to address the challenges of AI certification, with \citet{vanderlinde2022} focusing on the development of potential solutions and \citet{mahilraj2023} highlighting the issues of robustness, transparency, reliability, and safety in AI systems.
Beyond the previously mentioned ``black-box" nature of AI, where the internal workings and decision-making logic are often opaque \cite{castelvecchi2016}, two additional challenges are \emph{data dependency} and  \emph{continuous learning}.

\emph{Data dependency} is a critical issue for AI models. The quality of the data used for training, testing, and validation directly impacts the AI's performance \cite{Landgrebe.2022}.
Ensuring that datasets are robust, representative, and free of errors or biases is a complex task that diverges from traditional software validation methods, which do not typically rely on large datasets to function correctly.
\citet{picard2020} specifically address the need for dataset engineering in safety-critical systems, while \citet{budach2022} empirically explore the correlation between data quality dimensions and ML algorithm performance.
The lack of complete understanding of the nuanced interactions between AI models and their training data further complicates certification efforts.

The \emph{dynamic behavior} of AI systems that continue to learn post-deployment introduces an additional layer of unpredictability.
Such systems can evolve and adapt, potentially developing unforeseen and harmful behaviors that were not identified during the initial certification process.
This contrasts with traditional software, where behaviors can be tested and certified against a fixed set of specifications.
To overcome this issue, \citet{bakirtzis2022} propose a dynamic certification approach for autonomous systems, which involves iterative testing and revision of permissible use-context pairs. This approach allows for certification while learning, addressing the uncertainty and heterogeneity of deployment scenarios.
Regarding the dynamic behaviour of such AI systems, one of the main challenges is to establish certification processes that are flexible and robust enough to account for and monitor these changes over the lifetime of the AI system \cite{Stodt.2023}.




\subsection{XAI and its Role in Certification}

Regarding the previously described challenges of safe AI development, several publications propose XAI as a potential solution.
Some of them are presented in the following.

An intensely debated topic regarding safe AI is autonomous driving.
As such, several publications describe possible avenues of using XAI in that use case, describing it as a fundamental building block of autonomous driving systems \cite{kuznietsov2024explainable, Atakishiyev2021XAIautonomousDriving}.

In a more general setting, \citet{Brajovic.2023} describe a general framework for the documentation of AI (as precursor to certification), based on model cards \cite{mitchell.2019} and data cards \cite{pushkarna.2022}, including XAI as a potentially necessary part of development.
A similar notion is provided by \citet{Martinie.2021}, who view XAI as key to make AI in critical interactive systems transparent to users and certification stakeholders.
An application for these use cases is shown by \citet{Saraf.2020}, as they develop a proof of concept tool to generate local explanations for a trajectory anomaly detection model to demonstrate how XAI can help towards user acceptance and certification.

To be able to assess transparency in the safe development and certification, robust measures for XAI need to be developed and integrated into AI assessment \cite{Stodt.2023}.

Further evidence for the importance of explanations for safe AI can be found in recent standards, e.g. ISO 21448 (Road vehicles - Safety of the intended functionality).
This standard recommends analyzing the interpretability of ML software to increase its trustworthiness by showing that decisions are based on relevant data and not on artefacts.
As such, interpretability analysis can be included as pass/fail criteria into ML validation and/or test verification.

Some criticism is also levied against such approaches, as \citet{Landgrebe.2022} describe how the aim of XAI shifted away from the initial intentions of providing objective understanding of ML models.
As an alternative to XAI in safe AI development, they propose ``certified AI" via specification, realization and tests, including elements of ontology and formal logic within their AI approach.

While the utility of XAI in enhancing transparency is widely recognized, there is a notable gap in empirical research concerning its integration into the certification processes of AI systems. Most existing studies focus on theoretical frameworks or specific use case scenarios, with less emphasis on systematic, empirical evaluations of XAI's role in the broader certification processes \cite{Landgrebe.2022, Brajovic.2023}. This study aims to fill this gap by examining firsthand experiences of XAI in practice and evaluating its potential and limitations in the context of AI certification.


\section{XAI Taxonomy}
To contextualize the previously described publications and the interviews in the following chapters, a short overview over the current state of the art of XAI is presented here.

In general, different taxonomies for XAI exist, grouping XAI methods mostly by their general function, their type of result or via their underlying concepts \cite{Speith.2022}.
One of the most common distinction between methods is provided by their ``scope", e.g., whether XAI methods aim to explain a single decision (local), the functioning of an entire ML model (global) or the dataset of an ML task.

Some of the earliest---and most common---XAI methods are so-called feature-importance methods, either model-agnostic ones like LIME \cite{ribeiro2016lime} or SHAP \cite{lundberg2017shap}, or model-specific ones, mainly for computer vision tasks with neural networks, like Integrated Gradients \cite{sundararajan2017intgrad}.
They present their explanations as the ``importance" of features towards a decision, e.g., via highlighting input values of tabular data or as saliency maps (``heatmaps") that highlight (super-)pixels of images.

Because such explanations can provide ambiguous and thus, difficult to understand, information by only highlighting an area of an image without further information whether the form or the texture of some object is used, concept-based explanations were proposed \cite{Kim2017tcav}.
They try to decompose single decisions (or the general logic of an ML model) into human-understandable concepts, e.g., via concepts such as ``striped" or ``square-shaped".
In combination with such concepts, data-based explanations can be used, which show data instances similar to the input in question or relevant for the concept(s) used \cite{Achtibat2024crp}.

Another common explanation type are so-called ``counterfactual" explanations.
They provide information of the type ``If feature $X$ would have value $y$, the outcome would be different", e.g., if a user wants to know why they were not granted a loan.
By design, such explanations only provide a limited amount of information to prevent reverse-engineering of the model, but equip users with enough information to be able to challenge a decision or adapt accordingly to receive a different outcome \cite{Wachter2017}.

Another way to explain ML decisions is called ``mechanistic interpretability", a bottom-up approach which tries to decompose models into fine-granular explanations by taking their exact computations into account \cite{Bereska2024mechanistic}.
Corresponding explanations often entail specific neural circuits that are linked to specific behaviors or concepts (comparable to parts of \cite{Achtibat2024crp}).
Similarly concerned with exact computational behavior of ML models is the field of formal verification, where methods aim at formal or statistical guarantees for specific properties such as robustness against specific input perturbations \cite{Landers2023RLverification}.
While theoretically sound (and especially thought to be relevant for ML safety \cite{Landers2023RLverification}), such approaches often struggle with the computational complexity of neural networks for real-world applications.
Often not viewed as part of explainability itself, the closely related field of uncertainty quantification tries to provide ML decisions with uncertainty estimates, enabling users to spot potentially unsafe model decisions \cite{Abdar2021Uncertainty}.
Some authors also call for uncertainty (or ``confidence") estimates of explanations themselves to show whether a generated explanation should be trusted \cite{Nauta2023XAIsurvey,Fresz2024}.

Another approach is to adapt the model structure to inject previous knowledge about the data structure and to assist in explanation generation.
As indicated by their name, graph-based neural networks adapt the network structure to allow the interpretation as a graph, potentially improving data approximation and explainability \cite{Agarwal2023GNN}.
Neuro-symbolic approaches combine deep-learning approaches, e.g., for perception tasks \cite{Evans2021NeurosymbolicApperception}, with classical reasoning, to not rely on post-hoc explanations but to understand local decisions and the global logic of the resulting model \cite{Garcez2023Neurosymbolic}.

Summarizing, a lot of different approaches to explaining ML decisions and improving ML safety exist.
In this study, their shortcomings and potential in regard to AI certification---as viewed by XAI and certification experts---are evaluated.


\section{Methodology}

To survey the potential of XAI in general and in certification processes in particular, $15$ interviews were conducted.
In the following, the methodology for the interviews is described, including the participant profiles and the interview and coding process.

\subsection{Participant Profiles} \label{ssec:participant-profile}
To limit culture-specific influences, all participants either originated from Germany, Austria, or Switzerland or live there permanently.
Because of that, most interviews were conducted in German ($10$), and some in English ($5$).
Inclusion criteria for participants are knowledge about AI certification and about XAI, established by current projects or published works about at least one of these topics.
Since those requirements limit the number of potential interview candidates, purposive sampling \cite{Guest.2014} was used and participants were approached from the authors' individual networks.
An anonymized list of all participants can be found in Table~\ref{tab:participant-profile}.
Note that two further interviews were conducted, but due to a lack of expertise in either (X)AI or certification, they were not considered for further evaluation.
Since only a limited set of potential interviewees exist, further information about their background is omitted to keep them non-identifiable.
Where such information is relevant as background to certain statements, e.g., in Section~\ref{ssec:general-remarks}, the most limited amount necessary is given. 

\begin{table}[btp]
\centering
\begin{tabular}{ccc}
\hline
Identifier & Certification & (X)AI \\
\hline
P1         & 4                        & 4               \\
P2         & 4                        & 3               \\
P3         & 4                        & 3               \\
P4         & 4                        & 3               \\
P5         & 4                        & 3               \\
P6         & 4                        & 2               \\
P7         & 3                        & 4               \\
P8         & 3                        & 4               \\
P9         & 3                        & 4               \\
P10        & 3                        & 3               \\
P11        & 3                        & 3               \\
P12        & 3                        & 3               \\
P13        & 3                        & 2               \\
P14        & 2                        & 4               \\
P15        & 2                        & 3               \\ \hline
\end{tabular}
\caption{Relevant expertise of the interviewees. For (X)AI-expertise, the following conventions were used: 0 $=$ no expertise, 1 $=$ working expertise with AI, 2 $=$ working expertise with AI and experimenting with XAI, 3 $=$ extended XAI knowledge (without XAI being the focal point of the own work), 4 $=$ active research on XAI. Similar conventions were used for the certification expertise.}
\label{tab:participant-profile}
\end{table}

\subsection{Interview Process}
The interviews for this study were conducted from January to March of 2024 via Microsoft Teams.
After the agreement of participants to participate in the study, the general outline of the interviews was presented, including the following structure:
\begin{itemize}
    \item \textbf{Participant profile:} Participants were asked about their current position, their previous work and their current task relating to AI.
    \item \textbf{Use of (X)AI:} Participants were invited to explain their current aim and use of XAI, since challenges and the state of the art might differ between the aim and field of use.
    \item \textbf{XAI in certification:} After participants spoke about their experience with XAI in general, they were asked about specific challenges and requirements for XAI in the field of certification of AI.
    \item \textbf{Look into the future:} Since most of the previous questions focused on challenges in the field of XAI, participants were invited to share their thoughts and hopes regarding the future development of XAI.
\end{itemize}
For the full list of interview questions, see Appendix~\ref{a-sec:interview-guide}.

\subsection{Coding}
For the coding of the interviews, an inductive qualitative analysis based on \cite{Mayring.2019} was used.
Two coders reviewed three transcripts of the interviews, met for an intermediate coding workshop and finalized the coding.
During this process, potential ambiguities were marked and discussed in a final coding workshop.

\section{Results}

In this chapter, the results of the 15 expert interviews about the use of XAI in general and in certification are discussed.
At first, the practitioners' perspective of XAI is presented, followed by their opinions on the use of XAI for certification tasks.
The main results are also summarized in Table~\ref{tab:main-statements}.

Since the interviews were conducted in a semi-structured manner, interviewees could also remark on more general issues of XAI and AI certification.
Such statements, which are not commonly found in relevant literature, are presented in Section~\ref{ssec:general-remarks}.
For a more detailed breakdown of the individual interviews, see Table~\ref{tab:summary} in Appendix~\ref{a-sec:interview-summary}.

\begin{table*}[bt]
\centering
\begin{tabular}{|l|p{6cm}|p{6cm}|}
\hline
\textbf{} & \textbf{High Potential} & \textbf{Low Potential} \\
\hline
\textbf{XAI in general} & \begin{itemize}
    \item Communication between domain/AI experts
    \item Clear guidance on when to use which XAI method
\end{itemize} & \begin{itemize}
    \item Explaining to lay users
    \item Explaining in situations where the underlying processes are too complex or not well understood
\end{itemize} \\
\hline
\textbf{XAI in certification/safe AI} & \begin{itemize}
    \item Plausibility check of ML model by developers
    \item Discovery of Bias/Errors
    \item Improved Data Understanding
\end{itemize} & \begin{itemize}
    \item Assurances about AI safety
\end{itemize} \\
\hline
\textbf{Future of XAI} & \begin{itemize}
    \item Increased focus on user needs
    \item New explanation types (concept-based, mechanistic, multi-modal)
    \item Uncertainty quantification of (X)AI 
\end{itemize} & \begin{itemize}
    \item Comprehensive measurement of transparency/explainability
\end{itemize} \\
\hline
\textbf{Future of AI Certification} & \begin{itemize}
    \item XAI as an additional asset of certification processes
    \item Formal verification of safety-relevant AI properties
    \item New AI approaches (e.g., neuro-symbolic)
\end{itemize} & \begin{itemize}
    \item XAI as a comprehensive answer to AI certification
\end{itemize} \\
\hline
\end{tabular}
\caption{Summary of the main statements by the interviewees about the current state and future development of XAI.} \label{tab:main-statements}
\end{table*}


\subsection{Use of (X)AI}

\subsubsection{Aim of XAI Use}
The participants of the study unanimously thought of transparency or explainability as an important topic in safe AI development, which was to be expected since all of them are known to work on related topics (see Section~\ref{ssec:participant-profile}).
While they considered transparency and explainability as important, a common thread that emerged is that the integration of appropriate methods into standard development processes is still lacking in most cases.
This shortfall is partly due to the perceived lack of sufficient value-addition from explainability to warrant the necessary funding, especially when AI projects are externally commissioned.
When applied, the objectives of explainability methods are multifaceted and often abstract, encompassing aspects such as enhancing the public perception of projects, adhering to regulatory or customer requirements, detecting errors in ML systems, and facilitating internal communication about the capabilities and operation of ML methods across different departments (e.g., compliance and ethics checks).

\subsubsection{Choice of XAI Method}
With the aims of XAI use varying significantly, a clear framework for assessing the performance of XAI methods was not discernible from the interviews, also due to the wide range of XAI methods employed.
These span from neuro-symbolic ML systems, graph- and concept-based explanations to white-box models, and feature importance methods like SHAP and LIME.
The multitude of available methods and the ambiguity in evaluating the respective objectives make it challenging for practitioners to identify the most suitable method for a particular application.
Consequently, methods that are easy to implement and provide accessible information are often chosen.
Due to its open-source nature and ease of use, SHAP is commonly used, although the interviewees are aware of the criticisms levied at this method, e.g., by \citet{slack2020, kumar2020SHAPcriticism}, and thus sceptical of its performance and reliability.

\subsubsection{Experiences with XAI}
Despite the challenges described before, the interviewees reported successful applications of XAI procedures, particularly in identifying errors in existing ML systems, conducting plausibility checks on models during development, and enhancing data understanding.
However, it was also noted that current XAI methods are not well-suited for all use cases, with projects often failing due to common reasons.
These included the incomprehensibility of generated explanations to the target audience, lack of time for experts to interact with the explanations, and difficulty in verifying found correlations due to insufficient AI or domain expertise.
Further complicating the deployment of XAI is the unstable nature of some methods, leading to non-reproducible results, and the lack of comprehensive research on methods for specific data types like time series.

The use of XAI methods to foster trust among end-users, often highlighted in scientific literature, was viewed critically in many interviews.
The complexity of XAI methods effectively shifts the problem of an untrustworthy black-box (the ML system) to another black-box (the XAI method), the trustworthiness of which is also questioned due to the controversial nature of existing XAI methods.
This issue is exemplified by the disagreement problem \cite{krishna2022disagreement}, where different XAI methods provide different explanations for a single decision of an ML model, making it unclear what the ``true" explanation is.
Note that the black-box nature of the XAI method stems more from the lack of in-depth expertise about XAI than from a general incomprehensibility such as for the ML model.
Because of this, a potential solution mentioned by P2 to the trust issue created by the double black-box is the provisioning of training on AI and XAI for users.

\subsection{XAI in Certification}
Additional to XAI in development, interviewees were asked about their expectations and perceived challenges of XAI in AI certification.
Central to this discussion is the challenge of measuring ``appropriate" transparency and explainability in XAI methods (as demanded by the AI Act), a task that varies significantly depending on the specific purpose and function of the AI system in question.

\subsubsection{Use of XAI in certification}
Overall, two main groups of opinions about the use of XAI in certification can be distinguished:
From the perspective of some experts, the influence of XAI on the certification of AI systems is seen to be minor. This viewpoint stems from the existence of other regulatory measures like thresholds for certain performance metrics for AI systems or the belief that XAI methods, particularly in complex applications, are not and cannot be sufficiently robust or comprehensive.
In such applications, explanations generated by XAI methods themselves become too intricate, thus detracting from their utility.
This viewpoint is underpinned by the interviewees almost unanimously agreeing that explainability is not (an will not be) truly measurable, or will at least require user studies to do so.

In contrast, other experts---especially ones who successfully used XAI in the past---maintain that XAI has demonstrated its potential in improving ML models by identifying problems early in the development process.
As the main goal of safeguarding and certifying AI is to prevent potentially harmful defects, it is argued that XAI---even without quantitative performance metrics---helps towards that goal and should thus be part of AI certification.
Interviewees with this viewpoint often additionally pointed out that XAI could only be one of many tools for AI certification (additional to testing, formal verification etc.), providing only a small piece of evidence for certification.
Additionally, they emphasized that human inspectors remain integral to the certification process.

\subsection{Expectations for XAI}
Looking towards the future, the expectations and hopes associated with the development of XAI are diverse.
While the ideal of achieving complete, globally applicable explanations is largely seen as unattainable, some optimism persists around the evolution of new XAI approaches. These include concept-based, mechanistic, and neuro-symbolic methods, which are hoped to enable a new form of explanations elucidating the fundamental operation of ML models.

The interviewees also highlighted the necessity of user-centric and industry-focused approaches in order to fully realize the potential of XAI.
While XAI methods are seen as offering the capacity to detect errors in ML systems, and thus should ideally be integrated into the development processes, other methods are expected to be of higher importance for the certification landscape.
These include AI examination by alternate AI systems, formal verification of specific properties, and uncertainty quantification of AI decisions.
A major challenge for the explainability of AI systems is seen in the difference of new AI paradigms, as future Large Language Models (LLMs) might provide multi-modal inputs and outputs and explanations for e.g. time series or image data need to be fundamentally different than ones for other data types.

A recurring theme across discussions was the call for clear, definitive requirements for AI certification, such as specific metrics.
Without such clear guidance, the interviewees felt like companies would lack the available resources and information to ensure that their AI systems comply with the relevant transparency requirements, such as those in the AI Act.
This call for clear requirements is complemented by the advocacy for the use of simple, intrinsically interpretable AI solutions, wherever possible.
The use of AI in high-risk applications was considered inappropriate by some interviewees, emphasizing the need for caution and discretion in AI deployment.

\subsection{General Remarks about AI Certification and XAI} \label{ssec:general-remarks}
Additional to the more universal statements on XAI and XAI for certification, some interviewees also voiced concerns and opinions on specific topics.
Since these remarks are not commonly found throughout literature and believed by the authors to add interesting viewpoints to the discussion aimed at by this paper, they are presented in this section.
To clarify the distinction between the statement made and additional information provided to contextualize the statements during the writing of this paper, the initial statement is given in italic.
Note that these statements are not direct quotes.
Most of them were translated from German and edited for brevity and readability, as the direct quotes were spoken language and embedded in the context of the corresponding interview.

\subsubsection{Incorrect Evidence?}
P2 + P6: \emph{Certification so far examines whether evidence is in line with the requirements of standards and norms. There is no process in place to check whether this evidence is correct.}

In discussions with the experts, especially P2 and P6, it repeatedly emerged that the certification of AI presents new challenges compared to the existing processes for product certifications.
It was specifically pointed out that certification so far has been examining whether evidence provided by manufacturers conforms to the requirements of standards and norms.
It is assumed that this evidence is correct, meaning it corresponds with the used processes or the actually developed product.
If evidence is now to be generated using XAI, this assumption may not be correct.
Especially, since malicious actors may be able to generate almost arbitrary explanations for their ML systems, even with established methods \cite{slack2020,zhou2023}.
While providing incorrect evidence could be a possibility for classical systems as well, incorrect evidence might be provided by XAI without malicious intent of the stakeholders.
For such cases, responsibilities must be clarified: Do manufacturers guarantee the correctness of the evidence, which will hardly be possible with the current state of XAI, or must certifiers in the future consider the generation process of the evidence, which in turn requires in-depth knowledge regarding AI in general and XAI in particular?

P5: \emph{Until now, the ``Uniformity Hypothesis" and the ``Competent Programmer Hypothesis" were helpful pieces of building and certifying safety-critial systems.}

Similar to the point above, some previous assumptions might not hold for AI certification.
Usually, the here mentioned ``Uniformity Hypothesis" \cite{Sterling1969Uniformity} has been applied, as it describes that specific data points can be selected for tests, whose findings generalize across an equivalence class of similar data, and the ``Competent Programmer Hypothesis" \cite{DeMilloLS78CompetentProgrammer}, which postulates that safety-relevant software does not produce completely unpredictable errors because it was created by a competent programmer who can avoid errors that appear random (e.g., by buffer overflows or pointers, as commented by P5).
However, both assumptions are violated by the black-box nature of AI: For the generalizability of tests on certain data, equivalence classes are difficult to find and the decision-making process of an ML system was not explicitly programmed.
Due to the complexity of ML models, resulting errors may appear random.
Nevertheless, P5 noted that a good step towards safer AI is to document the assumptions made in AI and XAI, which is often not done for assumptions such as the ``Uniformity Hypothesis" and the ``Competent Programmer Hypothesis" in classical software development.
Regarding the criticism faced by some assumptions in XAI, P7 explicitly pointed out that scientific progress typically comes from challenging existing ideas.
In the field of XAI, this leads to a complexity that is difficult for practitioners to penetrate. Initially, XAI was proposed for the evaluation or testing of XAI itself, but now there are also metrics for the testing of XAI, and even metrics for evaluating those metrics \cite{Tomsett2019SanityChecks}.
Consequently, a goal of applied research could now be to make explicit recommendations on how to select XAI methods for specific use cases.

\subsubsection{Fundamental Changes in Certification}

P1: \emph{If AI is to be certified, there needs to be a discussion about a shift from value-based to utilitarism-based certification.} 

Due to the uncertainties in testing AI systems, an expert speculated that the culture of certification has to fundamentally change to accommodate AI: Existing certification is principally guided by societal values and norms.
For example, for the norm ``safety" of a technical system, a threshold can be defined, which can then be adhered to based on a detailed analysis of an overall system, for example through methods such as Failure Mode and Effect Analysis \cite{stamatis2003failure}.
If this value is not maintained, countermeasures must accordingly be taken from the development side.
A particularly well-known example of the traceability of ethical values in technical systems can be found in the field of autonomous driving, the so-called ``trolley problem".
In this scenario, an immediate choice must be made before an accident as to which involved individuals are subjected to a higher risk of severe injuries or potentially death.
For AI, however, such thresholds and ethical decisions are currently not sufficiently determinable.
Therefore, the expert suspected that the use of AI might need to be assessed more from utilitarian viewpoints, meaning ``If the use of AI is expected to result in fewer injuries or fatalities in traffic, then its use is sensible."

\subsubsection{Responsibility for Explainability Requirements}
P1 + P2: \emph{Explainability is more of a societal than technical topic, as such the standardization bodies are not well equipped to deal with it.}

To be able to certify a technical system, standards are used.
Tasked with creating such standards are organizations such as DIN (for Germany), CEN/CENELEC (European Committee for Electrotechnical Standardization), or ISO (International Organization for Standardization).
As these organizations commonly create technical standards, P1 and P2 argued that issues such as explainability and fundamental considerations like the compliance with and negotiation of ethical values (see above) should not be technical discussions, but rather socio-political debates.
It is also particularly noteworthy that while technical evaluation methods for XAI do exist, they were not considered to be effective by the majority of study participants, which suggests that purely technical standardization is unlikely to resolve the open questions surrounding the assessment and certification of AI.
P2 additionally noted that participants of standardization committees might lack the time to be well informed about topics as current as XAI (due to other obligations), thus resulting in standards that might not represent the current state of science.

\subsubsection{Fundamental Doubts about XAI}
P3 (with a background in neuroscience): \emph{For AI, the requirements are stricter than ever possible for humans. At best, XAI might provide justifications, while the only possible explanation for an AI system is its complete calculation from input to output.}

Around the topic of AI certification, there exists a discussion of whether AI should be subject to stricter requirements than humans doing the same task (as also touched on by \citet{Fresz2024}).
P3 extended this by linking explanations to the description of thought processes provided by \citet{Kahneman.2012}, dividing thought processes in system 1 thinking (fast, low effort, `intuitive') and system 2 thinking (slow, high effort, deliberate).
P3 argued that humans may justify their behavior upon request after the fact (system 2), but such justifications are not identical to the actual motives, especially for decisions that are often made intuitively (system 1).
They suggested that the same applies to XAI: XAI could produce a justification for ML behavior that is understandable to humans (system 2), but the true explanation could only be found within the computational chain of the ML system and, although fundamentally `transparent' (i.e., visible), not entirely understandable to humans due to the potentially huge number of calculations made by the ML system. 
It could be argued here that the ideal conception of XAI enables the computation chain to be summarized in such a way that a correct explanation is produced (e.g., via concepts), which provides users with insights into the ML behavior.

\subsubsection{Differences between Research and Practice}

P7 + P8 + P9: \emph{In research, cognitive load and interaction time with explanations are often not explicitly considered.}

Multiple participants criticized that in XAI research, the explicit experience and aims of domain experts are not considered enough.
They noted that XAI research seems to operate under the assumption that complete explanations should be generated in all circumstances. In contrast, domain experts, such as physicians, typically only require explanations in specific instances.

Furthermore, users are more likely to interact with and have a positive experience with explanations that serve to reduce the cognitive load associated with the task at hand.
The majority of users, in their daily routines, lack the time and cognitive resources to engage with overly complex explanations. This effectively undermines the core objective of XAI, which is to make AI more accessible.
To make explanations easier to understand, P1 mentioned that explanations need to be contextualized to fulfill their potential, which could potentially increase or decrease the cognitive load based on the specific implementation.
While interaction time and cognitive load are not commonly evaluated in XAI literature, there is some existing research that explores the idea of reducing the cognitive load of explanations \cite{herm2023cogLoad}.
This includes enforcing user interaction with explanations prior to presenting the ML recommendation \cite{miller2023XAIisDead}.

\subsubsection{New Paradigms for XAI}
P8: \emph{For the field of XAI, I am particularly optimistic about the feedback of XAI information into ML training.}

P8 identified the combination of the explanation process with the associated model improvement as particularly promising in the field of XAI.
So far, XAI has mostly been viewed unidirectionally---even if errors and biases can be identified in existing models, there is no simple way yet to intervene in the model or training data to correct existing problems.
A new paradigm (similar to the one proposed by \citet{pahde2023revealtorevise}) could offer the possibility to interact with explanations, correct them, and integrate these corrections back into the model training process.
Thus, insights gained from XAI could be efficiently used for the error correction of ML models.

\section{Discussion}
The conducted expert interviews illuminate potential pathways for the advancement of XAI, which will be explored in this section.
Additionally, constraints and limitations inherent in the study's design are addressed.




\subsection{Integration of Diverse Expert Perspectives}
Our research integrates insights from experts with dual expertise in XAI and certification. The diversity in expertise enriched the analysis, providing a well-rounded understanding of both the potential and limitations of XAI in AI certification processes. While our initial expectations anticipated these insights, the nuanced opinions offered by participants exceeded our predictions, underscoring the complex interplay between XAI capabilities and certification standards.
Due to the required expertise, only a limited amount of participants could be interviewed. However, due to the main opinions converging on either using XAI as an incomplete debugging tool or not at all in certification, no additional effect of more and similarly informed participants is to be expected.

\subsection{Critical Evaluation of XAI in Certification Contexts}
XAI's role in certification is pivotal yet constrained by several factors. First, while XAI can enhance transparency and facilitate error detection in AI models, its current capabilities do not fully satisfy the rigorous demands of certification standards which require better assurances of safety and reliability. This reveals a critical gap between the theoretical advantages of XAI and its practical utility in ensuring compliance with stringent certification protocols.

\subsection{XAI: From Debugging Tool to Certification Aid}
The dual utility of XAI as both a debugging tool and a potential certification aid presents a significant advancement in managing AI system complexities. As a debugging tool, XAI provides valuable insights into AI behavior, identifying biases and failure points. However, transitioning from debugging to a certification context requires XAI to offer more definitive guarantees (or at least information in the form of confidence estimates) about the correctness of explanations and AI systems’ behaviors and outcomes, a transition that is currently underdeveloped.

\subsection{Societal and Ethical Considerations}
The discourse around XAI goes beyond the technical boundaries and touches on the broader societal and ethical implications. Current certification frameworks primarily address technical compliance, but the integration of XAI requires a broader consideration of ethical standards and societal impacts. This requires a paradigm shift in certification, from purely technical evaluations to more holistic assessments that consider the societal implications of AI technologies.

\section{Summary}
As the field of AI continues to evolve, the adaptability of certification processes---and the role of XAI within these---will be paramount.
XAI is often touted as a potential solution to the black-box nature and thus, certification, of AI.
This notion was examined empirically in this paper via qualitative interviews with $15$ experts both in XAI and AI certification.
In these interviews, the current state of XAI was often viewed skeptically due to the known problems of such methods and the overall ease of use of more complex methods.
Despite that, the interviewees often came up with examples where they used XAI successfully.
In these, it showed that XAI is able to highlight errors in ML applications, while it does not seem well suited to provide simple and understandable explanations to end users or domain experts.
Based on the shortcomings of current XAI methods, our interviewees largely expect XAI to be at most a helpful asset in AI certification, but no comprehensive answer for the problems of AI certification.
The interviewees also highlighted further avenues for XAI research, especially into data types for which XAI methods are less common like time series and natural language and new explanation types like concept-based and multi-modal explanations.

Looking ahead, the integration of XAI into certification processes poses significant challenges and opportunities.
The evolving regulatory landscape, particularly with frameworks like the EU AI Act, will likely include explainability as a core component. However, the absence of standardized measures for assessing the sufficiency of explainability complicates this integration.
To be able to integrate XAI into certification processes, practitioners need clear guidance on which XAI method should be used when and future research must focus on developing robust, quantifiable metrics for (X)AI that align with certification standards and contribute effectively to the safety and reliability of AI systems.

\newpage
\appendix
\section{Interview Guide} \label{a-sec:interview-guide}

In the following, the interview guide for the interviews is provided.

\subsection{Interviewee Profile}
\begin{enumerate}
    \item What is your role in the company? 
    \item What is your background, jobwise and course of study?
    \item What’s the relation between AI and your company? Do you use it, test it, …?
\end{enumerate}

\subsection{(X)AI Use}
\begin{enumerate}
    \item What AI applications are used in your company? Are these self-developed or purchased? 
    \item To what extent do you consider transparency and explainability requirements in the development and deployment of AI applications? 
    \item Do you specifically use XAI methods to enhance transparency and explainability?
    \item 
        \begin{enumerate}
            \item If yes: \begin{itemize}
                \item Which methods are used?
                \item What is the goal of using them? 
                \item Do the current XAI methods assist in achieving this goal? 
                \item How is the achievement of the goal evaluated? Are there specific attributes that are particularly emphasized?
            \end{itemize}
            \item If no: Why not? What do you do instead? 
        \end{enumerate} 
    \item Are there any specific use cases or examples in your company where the use of XAI has been particularly challenging or successful? 
\end{enumerate}

\subsection{XAI in certification}
This part of the interview was started with a short explanation of the EU AI Act requiring ``sufficient" explainability of AI systems, followed by these questions:
\begin{enumerate}
    \item Do you think explainability/transparency is currently measurable enough to be assessed in a certification process? 
    \item What are the open questions regarding the measurability of appropriate explainability?
    \item Do you think XAI methods should be part of AI certification?
    \item What should XAI methods fulfill to be helpful in the certification process?
    \item In your opinion, does XAI allow fulfilling of transparency requirements (of the AI Act or other regulations)?
\end{enumerate}

\subsection{Outlook}
\begin{enumerate}
    \item Which new trends or technologies in XAI do you see as particularly promising? 
    \item How do you envision the future of XAI, especially in terms of ethical and regulatory aspects?
\end{enumerate}

\section{Interview results} \label{a-sec:interview-summary}
In Table~\ref{tab:summary}, all conducted interviews are summarized briefly, to provide a better overview over the statements given.
Note that the information given is kept general to keep the interviewees non-identifiable.

\begin{table*}
\begin{tabular}{|p{0.7cm}|p{0.8cm}|p{0.8cm}|p{4cm}|p{3cm}|p{4cm}|p{1.7cm}|}
\hline
Iden-tifier & Certifi-cation & (X)AI & Noteworthy failed/successful projects with XAI & Opinion on XAI State of the Art & Opinion on XAI in certification & Expected impact on certification \\
\hline\hline
P1 & 4 & 4 & Project failed due to explanations being too complex for users. Successful projects with explanations plus contextualisation. & Not yet where it should be. & Helpful asset, since errors can be found. Difficult to evaluate XAI with user studies due to individual differences in users. & Medium \\
\hline
P2 & 4 & 3 &  & Not yet where it should be. & Helpful asset. & Medium \\
\hline
P3 & 4 & 3 & Project showed new clusters in data, which were deemed sensible by domain experts. & ``True" Explanations will not be possible (see Section~\ref{ssec:general-remarks}). & Helpful asset for error detection, improvements of data knowledge. & Low/ Medium \\
\hline
P4 & 4 & 3 &  & No hope for the development of global explanations, overall not yet where it should be. & white-box models should be used and AI should not learn during deployment. XAI not really helpful for certification. & Low \\
\hline
P5 & 4 & 3 &  & Not yet where it should be. & Helpful asset. Assumptions in XAI methods should be documented (see Section~\ref{ssec:general-remarks}). & Medium \\
\hline
P6 & 4 & 2 &  & Not yet where it should be. & XAI only truly relevant when guarantees for (X)AI can be given. & Low \\
\hline
P7 & 3 & 4 & Project, where XAI showed errors in ML application for image data. & Not yet where it should be. & Helpful asset. & Medium \\
\hline
P8 & 3 & 4 &  & Not yet where it should be. & Helpful Asset. & Medium \\
\hline
P9 & 3 & 4 &  & Not yet where it should be. & Helpful Asset. & Medium \\
\hline
P10 & 3 & 3 & Failed to produce useful explanations for time series, successful for image data. & Not yet where it should be. & Hopes for formal verification/robustness analysis and performance metrics for AI certification. & Low \\
\hline
P11 & 3 & 3 & SHAP is used for internal communication (given enough experience). XAI fails due to too complex models/pipelines. Counterfactuals fail due to too many immutable/sensitive attributes. & Not yet where it should be. & Helpful asset. & Medium \\
\hline
P12 & 3 & 3 & Tested XAI methods could not detect frequency domain features for time series. & Not yet where it should be. & XAI would need clear guidelines, then it would be a helpful asset. & Low/ Medium \\
\hline
P13 & 3 & 2 &  & Not yet where it should be. & Helpful asset. & Low/ Medium \\
\hline
P14 & 2 & 4 & XAI showed bias in text application. & Good due to the theoretical guarantees of methods (especially IntGrad). & Human interpretation and access to model and data important to make XAI helpful asset. & Medium/ High \\
\hline
P15 & 2 & 3 & Explanations failed due to being too complex and fundamental connections in data not known. & Not yet where it should be. & Helpful asset (especially for fairness). & Medium \\
\hline
\end{tabular}
\caption{Tabular summary of the conducted interviews. For the (X)AI-
expertise, the following conventions were used: 0 = no expertise, 1 = working expertise with AI, 2 = working expertise with AI and experimenting with XAI, 3 = extended XAI knowledge (without XAI being the focal point of the own work), 4 = active research on XAI. Similar conventions were used for the certification expertise.} \label{tab:summary}
\end{table*}

\clearpage
\bibliography{main}

\end{document}